\documentclass{article}

\usepackage{epsfig}
\usepackage{amsmath}
\usepackage{amssymb}
\begin{document}
\title{Local computer model emulating the results of the Pan et al. experiment}

\author{Walter Philipp$^{1,2}$, Guillaume Adenier$^3$,\\Salvador Barraza-Lopez$^{2,4}$ and Karl
Hess$^{2,4}$\\ 1 Department of Statistics, UIUC\\ 725 South Wright
St., Champaign, IL 61820, USA\\2 Beckman Institute for Advanced
Science and Technology, UIUC\\ 405 North Mathews St., Urbana, IL
61801, USA\\3 School of Mathematics and Systems Engineering\\
Vejdes plats 7, V\"axj\"o University, 351 95 V\"axj\"o, Sweden\\4
Department of Physics, UIUC\\ 1110 West Green St., Urbana, IL
61801, USA}
\maketitle
\begin{abstract}
It is a widespread current belief that objective local models can
not explain the quantum optics experiment of Pan et al. By
presenting a model that operates on independent computers, we show
that this belief is unfounded. Three remote computers (Alice, Bob
and Claire), that never communicate with each other, send
measurement results to a fourth computer that is in charge of
collecting the data and computing correlations. The result
obtained by our local simulation is in better agreement with the
ideal quantum result than the Pan et al. experiment. We also show
that the local model presented by Pan et al. that can not explain
the quantum results contains inappropriate reasoning with profound
consequences for the possible results of any local model that uses
probability theory.
\end{abstract}



\section{Introduction}

The paper of Pan et al. \cite{pan} presents experimental results
using quantum optics and in addition a special version of an
objective local theory (POLT). For the actual experiment we refer
the reader to the original reference and describe here only POLT,
their objective local model, in detail. Three photons were
detected in coincidence with a fourth one (i.e. two entangled
photon pairs are used) after either a linear or circular
polarization measurement. For each photon $i = 1,2,3$ the authors
have introduced ``elements of reality $X_i$ with values $\pm 1$
for $H'(V')$ polarizations and $Y_i$ with values $\pm 1$ for
R(L)'' polarizations. These elements of reality are treated as
random variables because they in turn depend on another common
random variable $\Lambda$. $\Lambda$ may, for example represent
some hidden property of the two detected photon pairs and may have
any mathematical form (e.g. that of a vector labelling all four
photons). It is now common practice \cite{peres} to introduce a
counterfactual argument of the following kind. Given that a
measurement has been performed with one given setting e.g. the
$xyy$ experiment resulting in the measurement $X_1, Y_2, Y_3$, one
could have performed the measurement using another setting, say an
$yxy$ experiment or a $yyx$ experiment resulting in $Y_1, X_2,
Y_3$ or $Y_1, Y_2, X_3$, respectively, with the same $\lambda$.
Here $\lambda$ is the value that the random variable $\Lambda$ has
assumed in the actual measurement. Therefore, extending this
argument to all $X_i$ and $Y_i$, the authors claim that the
elements of reality $X_i$ and $Y_i$ satisfy the relations
\begin{equation}
Y_1\cdot{}Y_2\cdot{}X_3 = -1,\qquad{}Y_1\cdot{}X_2\cdot{}Y_3 =
-1,{\qquad}X_1\cdot{}Y_2\cdot{}Y_3 = -1 \label{sau4}
\end{equation}
and
\begin{equation}
{Y_i}\cdot{}{Y_i} = +1 \label{saufn1}
\end{equation}
and therefore by Eq. (\ref{sau4}) that
\begin{equation}
X_1\cdot{}X_2\cdot{}X_3 = -1 \label{sau5}
\end{equation}
whereas Quantum Mechanics predicts a $+1$ result for
Eq.(\ref{sau5}). This results represents a contradiction between
POLT and Quantum Mechanics.
 As has been analyzed elsewhere in more detail \cite{khr,hpnp},
this kind of reasoning can be seriously questioned. Key to the
understanding of the critique is the following. The counterfactual
argument that one could have measured with another setting and the
same $\lambda$ does, in our opinion, not yet cause any harm and
may be logically correct. However, to assume further that the
results of the real experiment can be understood in terms of and
actually do contain elements all with identical $\lambda$'s to
obtain Eqs. (\ref{sau4}-\ref{sau5}) requires several additional
mathematical assumptions.\\
One of these additional assumptions, that permeates all arguments
that start with conterfactual reasoning, is that all random
variables involved can be defined on a single probability space.
As outlined in greater detail in \cite{hessphilipp}, in order to
model an experiment one first needs to start with a one-to-one
correspondence between the experiments and the elements $\omega$
of the probability space (see Feller \cite{feller}). A random
variable $\chi$ assigns to each indecomposable event $\omega$ a
measurement $\chi(\omega)$, i.e., a value that the random variable
$\chi$ assumes at $\omega$. In order to algebraically manipulate
the random variables, such as in the conclusion that Eqs.
(\ref{sau4}) and (\ref{saufn1}) imply Eq. (\ref{sau5}) it is
necessary that all the random variables are defined on the same
probability space. Because in \cite{pan} the results of Eq.
(\ref{sau4}) originate from three different experiments, each of
them having a different configuration of measuring equipment, and
performed over three different time intervals, Eq. (\ref{sau4})
needs to be rewritten in the form
\begin{equation}\label{philipp1}
Y_1'\cdot{}Y_2'\cdot{}X_3' =
-1,\qquad{}Y_1''\cdot{}X_2''\cdot{}Y_3'' =
-1,{\qquad}X_1'''\cdot{}Y_2'''\cdot{}Y_3''' = -1
\end{equation}
It is only when we equate
\begin{align}\label{eq:equalities}
&Y_1'=Y_1'',\quad{}Y_2'=Y_2''',\quad{}Y_3''=Y_3'''
\end{align}
that Eq. (\ref{sau5}) can be written in the form
\begin{equation}
X_1'\cdot{}X_2''\cdot{}X_3''' =-1
\end{equation}
Our model exploits the fact that a \emph{fourth} experiment is
performed, in a separate time interval. We \emph{do not} identify
the product of the results of three independent experiments with
the result of this fourth experiment
\begin{equation}\label{eq:result}
X_1^*\cdot{}X_2^*\cdot{}X_3^* =+1
\end{equation}
 because equating:
\begin{align}\label{eq:equalit}
&Y_1'=Y_1'',\quad{}Y_2'=Y_2''',\quad{}Y_3''=Y_3''',\quad{}
X_1'''=X_1^*,\quad{} X_2''=X_2^*\quad{}\text{and } X_3'=X_3^*
\end{align}
means that the six random variables figuring in Eqs. (\ref{sau4})
and (\ref{sau5}) can all be defined on a single probability space.
This clearly contradicts Vorobev's theorem \cite{vorob}, which
states that this may not be possible since the four threefold
joint distributions form a closed loop. In particular, if in
addition to the above mentioned parameter $\Lambda$, time and
setting dependent equipment parameters $\Lambda_{\bf e}(t_m)$ are
in play, then it can be easily shown that Eq.(\ref{eq:equalities})
does not hold, yet one still has an objective local model (here
${\bf e}$ denotes equipment setting and $t_m$ the measuring time).
Thus in this case the six random variables figuring in Eqs.
(\ref{sau4}) and (\ref{sau5}) can not be defined on a single
probability space. The simple reason for this is that each
measurement time $t_m$ labels a different light cone in the sense
of relativity theory and therefore a different set of causal
influences. These in turn may then signify different experimental
conditions and the need for different probabilistic setups as
well.\footnote{We remark in this connection that any detailed
theory of the Pan et al. experiment that uses relativity will need
to introduce
four-vectors that in turn introduce time.}\\

 Equations (\ref{philipp1}) and (\ref{eq:result})
also imply that the promised ``all versus nothing'' (a single
experiment ruling out local realistic theories in favor of
standard quantum mechanics) is not corresponding to the real
experiment, because four distinct experiments need to be carried
out before any conclusions can be drawn. Needless to say, that in
order to obtain statistically meaningful results, a large number
of these four distinct experiments will have to be performed. The
``all versus nothing" is based on the inappropriate assumption
that the results of the real experiment can be understood and
actually do contain elements all with the identical $\lambda$'s to
obtain Eqs. (\ref{sau4}-\ref{sau5}), as discussed above.

In the following we present a simple model that goes beyond POLT
by introducing time related variables. Our model is objective
local because it can be implemented on independent computers. We
are aware of the fact that the model is not the most general and
does not prove the existence of $\Lambda_{\bf e}(t_m)$. It also
does not provide any finality to the discussions surrounding the
theorem of Bell and other related experiments. However, we mantain
that oversimplified models such as POLT, (that employ only one
$\Lambda$ and one single probability space for four distinct
experiments) are inadequate for any serious argumentation designed
to prove non-locality in nature. Therefore, if one wants to make
statements about objective locality out of certain experimental
results, it will also be necessary to improve in the future the
actual experiments so that models such as the one presented below
can be ruled out.
\section{Description of the model}

We start our extended model involving independent computers by
introducing the following functions \cite{hessphilipp}. Let
\begin{equation}
r_k(t) = sign[sin({2^k} \pi t)] \text{  for  }t>0 \label{sauf3}
\end{equation}
denote the $k$-th Rademacher function. Note that $r_k$ has period
$2^{-(k - 1)}$. The following table is to be implemented in the
network simulation.\\

\begin{table}[ht]
\begin{center}
\caption{Functions for the local computers that simulate the
results of the Pan experiment}
   \begin{tabular}{|l||c|c|c|c|}\hline
  & $yyx,t_0<t<t_1$ & $yxy,t_2<t<t_3$ & $xyy,t_4<t<t_5$ &
$xxx,t_6<t<t_7$
\\ \hline
     Comp1 & $Y_1'=-r_1$ & $Y_1''=-r_1$ & $X_1'''=r_2\cdot{}r_3$ & $X_1^*=r_2\cdot{}r_3$\\
\hline
     Comp2 & $Y_2'=r_2$ & $X_2''=r_1\cdot{}r_3$ & $Y_2'''=r_2$ & $X_2^*=r_1\cdot{}r_3$\\
\hline
     Comp3 & $X_3'=r_1\cdot{}r_2$ & $Y_3''=r_3$ & $Y_3'''=-r_3$ & $X_3^*=r_1\cdot{}r_2$\\
\hline
  \end{tabular}
 \label{table:model}
\end{center}
\end{table}

Here  $t_1 - t_0$ is the length of time the $yyx$ experiment is
running, $t_2 - t_1$ is the length of time it takes the
experimenters to switch the experimental set-up from an $yyx$
experiment to an $yxy$ experiment. $t_3 - t_2$ is the length of
time the $yxy$ experiment is running and $t_4 - t_3$ again the
time to switch and so forth as described in Table
\ref{table:model}. Each of the three equations in Eq.
(\ref{philipp1}) holds on the entire time interval where they are
defined. Moreover, we have
\begin{equation}
X_1^*\cdot{}X_2^*\cdot{}X_3^* = +1 \label{saucorr5}
\end{equation}
instead of Eq. (\ref{sau5}), if we mimic at a later time the $xxx$
experiment according to the last column in Table
\ref{table:model}. Furthermore, each X and each Y equals +1 or
$-1$ half of the time. The essential point here is, of course,
that for a given time in which a given experiment e.g. $yxy$, is
being performed, we can assume that equipment parameters are such
that $Y$ may be described by a certain Rademacher function, e.g.
$Y_3 = r_3$, while at a later time, for which the $xyy$ experiment
is performed we may have $Y_3 = -r_3$. (Here we have made use of
the fact that in the Pan et al. experiment the settings e.g. $yyx$
are set and used for a longer period of time so that communication
between the stations is possible by sub-light velocities.) For a
given setting, the outcomes of the various experiments are, of
course, only ``known'' at a given detector, not at the others.
This is so because, as will be further described in the
implementation, each of the players have only the part of the
table that corresponds to them, while the Host only determines
when the experiments start, and he does not have any knowledge of
the entries on Table \ref{table:model}. Only the choice of
measurement time, which is random, determines the outcomes
together with the Rademacher functions that are characteristic for
a given setting. Of course the three Rademacher functions in Table
\ref{table:model} can be replaced by three other Rademacher
functions with arbitrarily large but different subscripts if
faster fluctuation between +1 and $-1$ is desired.

\section{Implementation of the model}

The networking program works as follows. We make use of a Visual
Basic template available online, and engineer it such that three
computers (Alice, Bob and Claire) are connected to a fourth
computer (Host). The Host first listens from its ethernet port
until the three computers are connected. The Host is able to
determine whether or not all the three remaining computers are
reachable with the given domain. This is done by:
\begin{itemize}
\item{}Pressing the \emph{Port Settings} button
(by default set to \emph{First Local Port} = 600, \emph{First
Remote Port} = 700, and \emph{Maximum Number of Connections} = 3).
\item{}Pressing the \emph{Host} button.
\end{itemize}
 The three players connect to the
remote Host whose IP number should be entered by each player in
their \emph{Connect To} window, with the settings
\begin{itemize}\item{}
\emph{First Local Port} = 600. \item{} \emph{First Remote Port}=
701 for Alice, 702 for Bob and 703 for Claire.\end{itemize} When
the three players are connected, the Host can see a new button,
which enables him to
\begin{itemize}\item{}
\emph{Launch the Simulation}. \end{itemize} By pressing it, it
tells Alice, Bob and Claire, to set a common time label, and to
start sending data according to Table \ref{table:model} at certain
time intervals.

At a given time $t$, the player at computer $i$ will make either a
$x$ experiment, with a measurement result denoted $X_{i,t}$, or a
$y$ experiment, with a measurement result denoted $Y_{i,t}$, the
result of each measurement being either $+1$ or $-1$, according to
Table \ref{table:model}. The players perform successively four
sequences of ``experiments'': a sequence of $yyx$ experiments
(Alice and Bob perform a $y$ measurement, while Claire performs a
$x$ measurement), a sequence of $yxy$ experiments, a third
sequence of $xyy$ experiments and finally a sequence of $xxx$
experiments. Each player sends his/her label ($x$ or $y$)
according to his/her performed experiment, together with a
measurement result to the Host, in charge of collecting the data
and processing the product of the three measurement results sent
by the players. The code was written so that, at a given time,
each player can find in the code only the relevant part of Table
\ref{table:model}, while this table is unaccessible to the Host.
Very importantly, during the process, the players never
communicate with each other, only \emph{unidirectionally} to the
Host. This is to say, they are only
sending data through their ethernet cards.\\

This networking simulation exhibits a perfect agreement with the
ideal quantum results. The data sent to the Host by the three
players Alice, Bob, and Claire successively yields the following:
\begin{equation}
Y_1'\cdot{}Y_2'\cdot{}X_3' =
-1,\qquad{}Y_1''\cdot{}X_2''\cdot{}Y_3'' =
-1,\qquad{}X_1'''\cdot{}Y_2'''\cdot{}Y_3''' = -1
\end{equation}
and finally,
\begin{equation}
X_1^*\cdot{}X_2^*\cdot{}X_3^* = +1,
\end{equation}
in accordance with Eqs. (\ref{philipp1}) and (\ref{eq:result}).\\

The detailed program can be obtained from us by e-mail upon
request at the following electronic address:
guillaume.adenier@msi.vxu.se.
\section{Conclusion}

We have presented a game that obeys all the rules of locality that
are obeyed in the actual experiment of \cite{pan} and that results
in the quantum mechanical correlations. With respect to the
latter, our simulation outperforms the experimental results of
Pan, Bouwmeester, Daniell, Weinfurter and Zeilinger \cite{pan},
and therefore clearly invalidates their claim that objective local
parameters cannot explain their experimental results. Future
experiments and models need to consider therefore the effects of
time and setting dependent equipment parameters.

\section{Acknowledgements}
S.B.-L. acknowledges partial funding from CONACyT (Mexico). He
also is grateful to the School of Mathematics and Systems
Engineering, University of V\"axj\"o, for their hospitality during
his stay. Support of the Office of Naval Research
(N00014-98-1-0604) is gratefully acknowledged.

\end{document}